\documentclass[10pt,conference]{IEEEtran}
\IEEEoverridecommandlockouts
\usepackage{cite}
\usepackage{amsmath,amssymb,amsfonts, bm}
\usepackage[ruled]{algorithm2e}
\SetKwRepeat{Do}{do}{while}
\usepackage{graphicx}
\usepackage{textcomp}
\usepackage{xcolor}
\def\BibTeX{{\rm B\kern-.05em{\sc i\kern-.025em b}\kern-.08em
    T\kern-.1667em\lower.7ex\hbox{E}\kern-.125emX}}

\definecolor{ao(english)}{rgb}{0.0, 0.5, 0.0}

\usepackage{multirow}
\usepackage[normalem]{ulem}
\useunder{\uline}{\ul}{}
\usepackage{subcaption}
\usepackage{graphicx}
\begin{document}
          
\title{QuMoS: A Framework for Preserving Security of Quantum Machine Learning Model\\


}

\author{\IEEEauthorblockN{
Zhepeng Wang\textsuperscript{\dag, \S},
Jinyang Li\textsuperscript{\dag}, 
Zhirui Hu\textsuperscript{\dag}, 
Blake Gage\textsuperscript{\ddag},
Elizabeth Iwasawa\textsuperscript{\ddag},
Weiwen Jiang\textsuperscript{\dag, \S}}

\IEEEauthorblockA{\textsuperscript{\dag}George Mason University, Department of Electrical and Computer Engineering, VA, USA.\\
\textsuperscript{\S}George Mason University, Quantum Science and Engineering Center, VA, USA.\\
\textsuperscript{\ddag}Leidos, VA, USA.\\
\{zwang48, wjiang8\}@gmu.edu
\vspace{-0.15in}}
}


\maketitle

\begin{abstract}

Security has always been a critical issue in machine learning (ML) applications.
Due to the high cost of model training --- such as collecting relevant samples, labeling data, and consuming computing power --- model-stealing attack is one of the most fundamental but vitally important issues. 
When it comes to quantum computing, such a quantum machine learning (QML) model-stealing attack also exists and is even more severe because the traditional encryption method, such as homomorphic encryption can hardly be directly applied to quantum computation.
On the other hand, due to the limited quantum computing resources, the monetary cost of training QML model can be even higher than classical ones in the near term.
Therefore, 
a well-tuned QML model developed by a third-party company can be delegated to a quantum cloud provider as a service to be used by ordinary users. In this case, the QML model is likely to be leaked if the cloud provider is under attack. 
To address such a problem, we propose a novel framework, namely QuMoS, to preserve model security.
Instead of applying encryption algorithms, we propose to divide the complete QML model into multiple parts and distribute them to multiple physically isolated quantum cloud providers for execution.
As such, even if the adversary in a single provider can obtain a partial model, it does not have sufficient information to retrieve the complete model.
Although promising, we observed that an arbitrary model design under distributed settings cannot provide model security.
We further developed a reinforcement learning-based security engine, which can automatically optimize the model design under the distributed setting, such that a good trade-off between model performance and security can be made.
Experimental results on four datasets show that the model design proposed by QuMoS can achieve a close accuracy to the model designed with neural architecture search under centralized settings while providing the highest security than the baselines. 

\end{abstract}
\begin{IEEEkeywords}
Cloud Quantum Computing; Quantum Machine Learning; Quantum Model Security.
\end{IEEEkeywords}

\section{Introduction}
\label{sec:intro}
\begin{figure*}[htbp]
    \centering
    \includegraphics[width=1\linewidth]{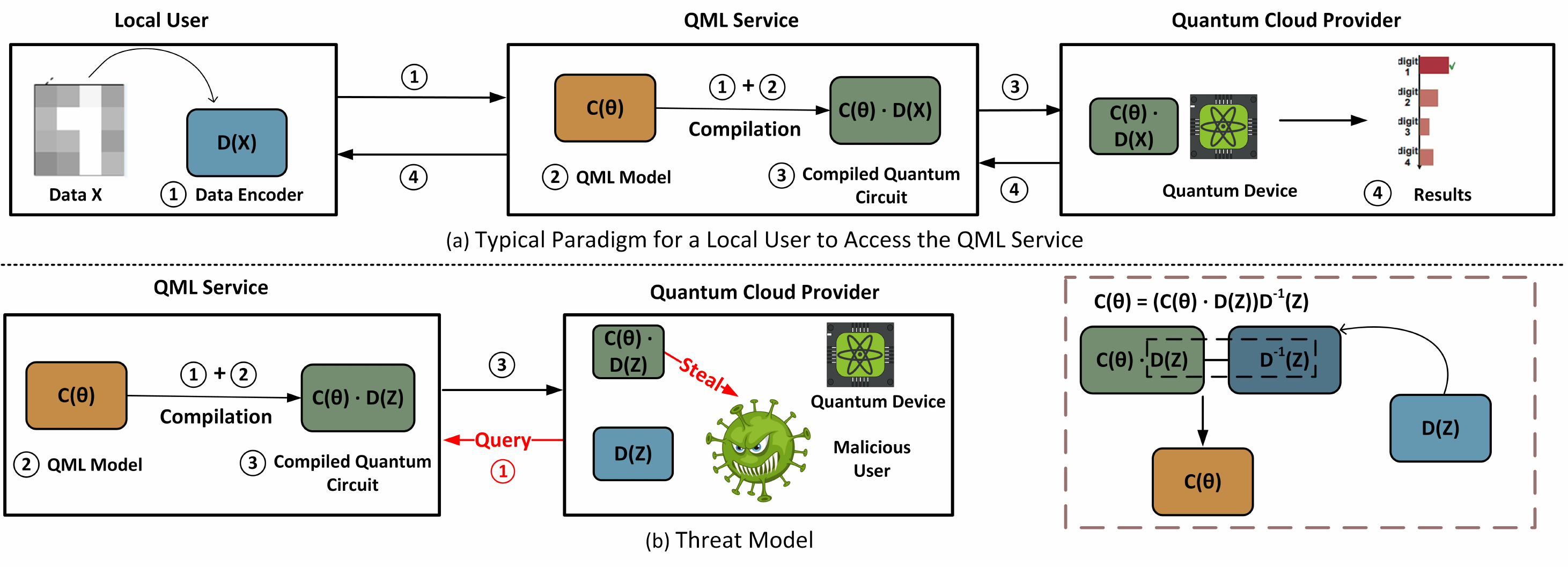}
    \caption{Threat model to steal the model in QML service}
    \label{fig: threat model}
\end{figure*}
Along with Google's first claim of quantum supremacy \cite{arute2019quantum}, researchers have been exploring varied quantum algorithms to exploit the superior computation power of quantum computers~\cite{shor1999polynomial, liang2022hybrid, liang2022pan, grover1996fast, kandala2017hardware, cao2019quantum}. In the meantime, the high capability of machine learning to process a large amount of data has made it widely used in a variety of applications such as image classification~\cite{huang2017densely, he2016deep, zhang2022toward, wu2022decentralized, wusynthetic, wu2021enabling}, natural language processing~\cite{vaswani2017attention, devlin2018bert, brown2020language, peng2022length, xu2023neighborhood} and medical diagnosis~\cite{ronneberger2015u, wu2021federated, wu2021federated_2, wu2022distributed, kan2022brain, kan2022fbnetgen}. Therefore, quantum machine learning (QML) is regarded as one of the most promising applications for its wide range of potential applications~\cite{cerezo2022challenges, liang2022variational, zhong2020quantum, peruzzo2014variational, rebentrost2014quantum, lloyd2013quantum, farhi2014quantum} and its potential to achieve an exponential speedup for model execution~\cite{liu2021rigorous, jiang2020co}, without the need to modify the original model like pruning~\cite{han2015deep, he2017channel, peng2022towards, wang2021lightweight} and quantization~\cite{krishnamoorthi2018quantizing, wang2019haq, wu2020intermittent, peng2021binary} in the machine learning on classical computers.



Although promising, in the near term Noisy Intermediate-Scale Quantum (NISQ) era, there are a lot of challenges.
One major challenge is the limited computing access: Unlike classical computing resources that almost everyone can easily access, today's quantum computing is mainly remotely accessed via the cloud services provided by quantum cloud providers, such as IBM, AWS, Google, Azure, etc, known as Quantum-as-a-Service (QaaS). Moreover, although there exist many works to accelerate the training process and reduce training costs for classical machine learning models~\cite{huang2022dynamic, bao2020fast, bao2022accelerated, bao2022doubly, bao2019efficient, wu2020enabling}, they cannot be directly used for QML models due to the intrinsic differences in the computer architecture. As a result, the training of a QML model is not free but time-consuming and expensive. From the AWS Pricing calculator, to train on the full MNIST dataset with 5 qubits requires over \$50K cost.
With such a large cost, it is impractical for ordinary users to train the QML model by themselves, and a better way is to utilize the existing well-tuned QML services provided by third-party companies.
As such, users only need to upload their data to the QML service, and the service will return the results to the users after the execution of their QML model on the data from users. 

Such a workflow looks practical; however, there exists a potential risk: the model from third-party companies obtained with a large cost has a high risk of leakage.
More specifically, since the service providers also need to execute QML model via quantum cloud providers, if  the quantum cloud providers are untrustworthy, the QML model could be easily decoupled from the quantum circuit. A simple threat model will be introduced in Section~\ref{sec:threat model}.

To defend against such attacks, we propose QuMoS, a framework for preserving the security of QML model by dividing the model into multiple parts and distributing them to multiple quantum cloud providers for execution. In this way, a single untrustworthy attacker can only get parts of the complete QML model. 
Although preserving security via distribution~\cite{shen2022distributed} has been studied in classical computing platforms, when it comes to QML, there exist a lot of challenges and open questions to be addressed.

Firstly, although classical distributed learning has been demonstrated to be effective, it is unknown whether the QML model can perform well under the distributed settings since the noise in quantum computers is usually much higher than in classical computers and the noise of quantum computers on different quantum cloud providers are usually different.
What's more, it is unclear how distributing QML model can provide a good trade-off between accuracy and security.
And we observed that an arbitrary design with a simple system layout (i.e., the topology of the computing node) will lead to low security, that is, the partial model obtained in a single quantum cloud provider can still achieve a good performance. The detailed results of a case study are shown in Section~\ref{subsec:motivation}.

In this paper, to the best of our knowledge, we propose the first quantum model security persevering framework, namely QuMoS, to address the problems mentioned above.
With the objective of maximizing security without compromising model accuracy, QuMoS develop a reinforcement learning-based  approach to automatically determine system layout, search for the best quantum architecture of a computing node, and assign the node to a quantum cloud provider.


The main contributions of this paper are listed as follows.
\begin{itemize}
    \item We reveal the security risk in quantum machine learning running at cloud-based quantum computing.
    \item To overcome the security issue, we propose the design of QuMoS, which preserve the security of the QML model by distributing and executing it on multiple quantum cloud providers.
    \item We further develop a security engine in QuMoS, which could find the best design of QML model and its configuration to distributed quantum computers, such that the performance of model is maintained in diverse noise environments across quantum cloud providers while preserving the security of model.
\end{itemize}

Experimental results on four datasets show that the design found by QuMoS can have close accuracy to models designed with neural architecture search under centralized settings, with even $0.67\%$ improvements on MNIST-4. Moreover, the security of model from QuMoS can consistently outperform the baselines on four datasets.


The rest of this paper is organized as follows. Section~\ref{sec:threat model} introduces the threat model. Then, in Section~\ref{sec:design}, the design of QuMoS is illustrated. Section~\ref{sec:search} further introduces the security engine of QuMoS. And the evaluation of QuMoS is conducted in Section~\ref{sec:experiment}. Section~\ref{sec:related work} outlines the related works of this paper, And Section~\ref{sec:conclusion} remarks the conclusion.

\section{Threat Model}\label{sec:threat model}



Fig.~\ref{fig: threat model} (a) shows the typical paradigm for a local user to access the QML service provided by a third-party company. More specifically, the user should encode its data $X$ into a quantum circuit (i.e., data encoder $D(X)$) and then send it to the QML service. The QML service will integrate $D(X)$ with its well-tuned QML model $C(\bm{\theta})$ and optimize it with a quantum compiler. The compiled quantum circuit $C(\bm{\theta})\cdot D(X)$ will then be sent to the quantum cloud provider for execution. And the results will be sent back to the local user through the QML service.

Fig.~\ref{fig: threat model} (b) shows a threat model targeting $C(\bm{\theta})$ shown in (a). Assume that the quantum cloud provider is untrustworthy, then it could send arbitrary data $D(Z)$ to query the QML service, pretending as an ordinary user. The QML service will send the compiled quantum circuit $C(\bm{\theta})\cdot D(Z)$ to the quantum cloud provider, which can thus be stolen by the attacker on it. Since the quantum circuit is reversible, the attacker could build the reverse quantum circuit $D^{-1}(Z)$ based on $D(Z)$. More specifically, by placing $C(\bm{\theta})\cdot D(Z)$ together with $D^{-1}(Z)$, $D^{-1}(Z)$ and $D(Z)$ will cancel out with each other. Therefore, the attacker can get the original $C(\bm{\theta})$.


\section{Design of QuMoS}
\label{sec:design}
\begin{figure*}[htbp]
    \centering
    \includegraphics[width=1\linewidth]{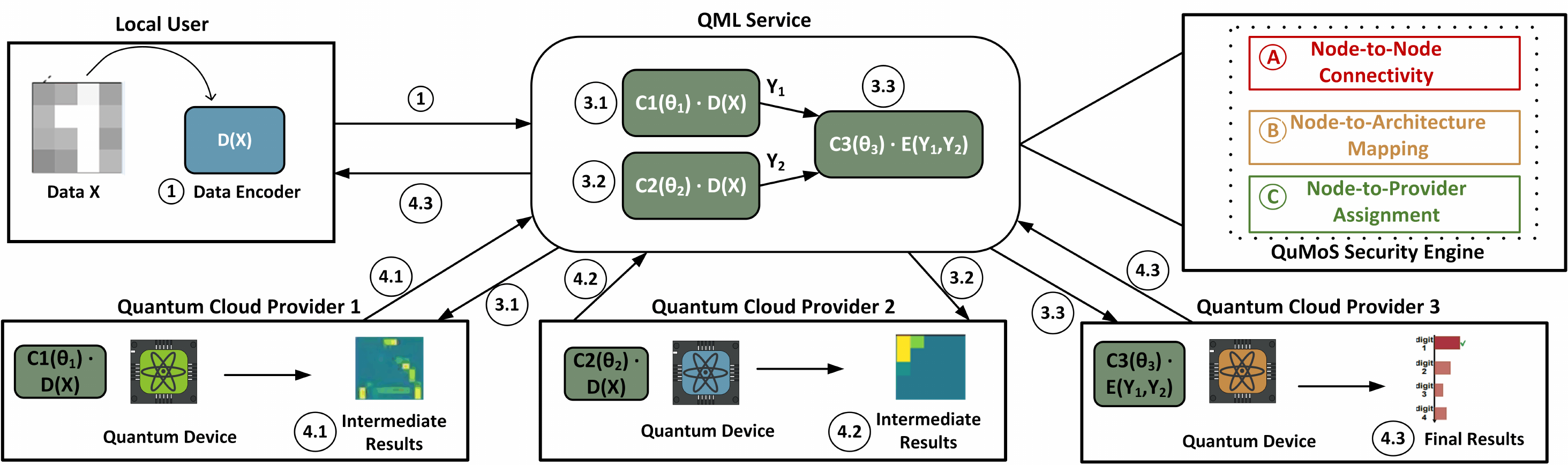}
    \caption{Design overview of QuMoS}
    \label{fig: design overview}
\end{figure*}

To defend against the attack shown in Section~\ref{sec:threat model}, we propose QuMoS, a framework to preserve the security of the QML model, which utilizes multiple quantum cloud providers to deploy and execute the QML model. In this section, we will illustrate the details of the design of QuMoS.

\subsection{Design Overview}~\label{subsec:design overview}
Fig.~\ref{fig: design overview} shows the design overview of QuMoS. In QuMoS, the QML model consists of multiple blocks. And each block will be distributed to the quantum device on a specified quantum cloud provider for execution. In this case, each provider only has partial information about the complete QML model. Therefore, a single untrustworthy provider cannot steal the complete quantum circuit of the QML model. To design such kind of QML model, QuMoS proposes a security engine to search for the best candidate. More specifically, by taking each block as a computing node, we can get a directed acyclic graph (DAG) corresponding to the design space of the QML model. Each node within the DAG has three basic properties, i.e., node-to-node connectivity, node-to-architecture mapping, and node-to-provider assignment, which will be defined in Section~\ref{subsec:properties}. The  security engine will then leverage reinforcement learning to decide the three basic properties for each node, which will be illustrated in Section~\ref{sec:search}. 

For the example in Fig.~\ref{fig: design overview}, the QuMoS security engine generates a QML model with three computing nodes, where the data $D(X)$ will be fed into the two head computing nodes $C1(\bm{\theta_{1}})$ and $C2(\bm{\theta_{2}})$ for processing. Then the generated intermediate results $Y1$ and $Y2$ will be utilized by the assembled node $C3(\bm{\theta_{3}})$, which will produce the final results. When executing such a QML model, the security engine decides to distribute each computing node to a different quantum cloud provider for security, which ensures that the arbitrary quantum cloud provider can only get part of the complete QML model.

\subsection{Definition of the Properties of Computing Node}~\label{subsec:properties}
\begin{figure*}[t]
    \centering
    \includegraphics[width=1\linewidth]{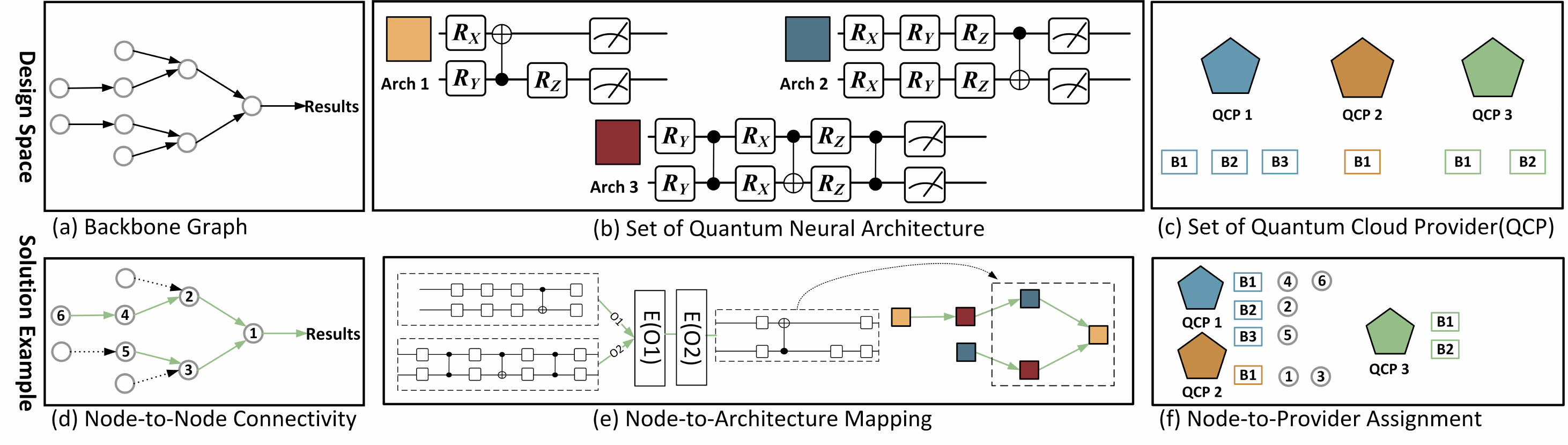}
    \caption{Basic properties of a computing node in QuMoS: (a)-(c) defines the design space. (d)-(f) shows an example solution to (a)-(c), respectively, Note that the basic properties of each computing node are annotated in (d)-(f).}
    \label{fig: design component}
\end{figure*}

In this part, we will define the three basic properties of computing nodes in QuMoS, as shown in Fig.~\ref{fig: design component}. 

\subsubsection{Node-to-Node Connectivity} It decides how each computing node will connect to other nodes in the QML model. To make such decisions, a backbone graph should be provided. And the subgraph sampled from the backbone graph can be served as the layout of the QML model, where the edge between the nodes represents a connection and the arrow represents the direction of dataflow. Based on the subgraph, the connectivity to the other nodes for a given node is determined. Fig.~\ref{fig: design component} (a) shows an example of the backbone graph, which has 9 optional computing nodes. Based on it, a subgraph with 6 nodes is sampled and used as the layout for the QML model, shown in Fig.~\ref{fig: design component} (d). Taking node 4 in the sampled subgraph as an example, we can know that node 4 is connected to node 6 and node 2 and it receives the data from node 6 for processing while sending the processed data to node 2. 

\subsubsection{Node-to-Architecture Mapping} To get a QML model with good performance, the design of the quantum neural architecture for each computing node is important, which is defined as the node-to-architecture mapping. As shown in Fig.~\ref{fig: design component} (b), a set of quantum neural architecture is provided. In Fig.~\ref{fig: design component}(e), each node will select one candidate from the set in (b) to be its architecture. Note that when the input of a given node is from the outputs of multiple nodes, the outputs will first be encoded with angle encoding~\cite{wang2022quantumnat} separately, and the corresponding encoding circuits will be stacked sequentially to serve as the inputs to the node. For instance, in Fig.~\ref{fig: design component} (e), the output $O1$ from node 2 and the output $O2$ from node 3 are encoded as the circuit $E(O1)$ and $E(O2)$, respectively. $E(O1)$ and $E(O2)$ are then stacked and followed by the circuit of node 1.

\subsubsection{Node-to-Provider Assignment} To preserve the security of the QML model, it is important to assign the nodes to quantum devices in different quantum cloud providers appropriately. And this assignment is defined as the node-to-provider assignment. As shown in Fig.~\ref{fig: design component} (c), a set of quantum cloud providers (QCPs) is given, where each available provider could have multiple quantum devices. For example, quantum cloud provider 1 (QCP 1) has three quantum devices, i.e., B1, B2 and B3. Given the set of quantum cloud providers, each computing node will thus be assigned to a quantum device of a specific quantum cloud provider, which is shown in Fig.~\ref{fig: design component} (f). For instance, node 4 is assigned to the quantum device B1 on QCP 1 for execution.

\subsection{Security Measurement}

For a given QML model $M$ in QuMoS, whose basic properties have been decided by the security engine, we need a method to measure its security. In QuMoS, there are three steps to take for the security measurement of $M$, which are listed as follows. 
\begin{enumerate}
    \item Find all of the security submodels within $M$. A security submodel $sm$ is defined as a QML model consisting of a set of nodes that are connected in $M$ and also assigned to the same quantum cloud provider. 
    \item Evaluate all of the security submodels on the dataset and get the corresponding accuracies.
    \item Calculate the heuristic security metric $SecMec(M)$, which is defined as follows.
    \begin{equation}
        SecMec(M) = 1 - \frac{\max_{q \in QS, sm \in SMS} ACC(sm, q)}{ACC(M)}~\label{eq:secmec}
    \end{equation}
\end{enumerate}

Where $QS$ is the set of all the utilized quantum devices and $SMS$ is the set of all the security submodels. $ACC(sm, q)$ denotes the accuracy of security submodel $sm$ on quantum device $q$, while $ACC(M)$ is the accuracy of the complete model $M$, which is evaluated in advance. The larger $SecMec(M)$ is, the higher the security of the QML model $M$ is.

Algorithm~\ref{Alg:Secsubmodels} shows the details to find all of the security submodels given a QML model $M$. The input to the algorithm is a directed acyclic graph (DAG) $G$ extracted from $M$, where each node has an attribute to specify the QCP it is assigned to. The algorithm has three stages. In stage 1, we build a hashmap $SG_{QCP}$ recording the subgraph for each QCP, where each subgraph contains all the nodes assigned to a specific QCP. In stage 2, we first need to ignore the direction of edges in each subgraph and thus take each subgraph as an undirected subgraph. For each subgraph, we can find its related connected components (i.e., security submodels) using the breadth-first search (BFS) algorithm. And the results are saved in $CC_{QCP}$. In stage 3, we need to reconsider the direction of the edges for each security submodel. More specifically, we will identify the starting nodes and end nodes within each submodel according to the indegree and outdegree of a node. When using the security submodel for inference, the data will be fed into the starting nodes while the results will be collected from the end nodes. The dataflow through the submodels will follow the topological order of the corresponding subgraph. The output of the algorithm is a hashmap $smap$ containing the information of all the security submodels within $M$. Since the time complexity of each stage is $O(N)$, where $N$ is the total number of nodes in $M$, we can conclude that the time complexity of the algorithm is $O(N)$.
\begin{algorithm}[!t]
\LinesNumbered 
\caption{Algorithm to find security submodels}
\label{Alg:Secsubmodels}

\KwIn{Directed acyclic graph (DAG) $G$ of a QML model $M$}
\KwOut{Hashmap $smap$ of security submodels}
\BlankLine

/*\texttt{\footnotesize Stage 1: Divide the graph into subgraphs, where each corresponds to a QCP}*/\\
$SG_{QCP} \leftarrow $ \textbf{HashmapInit()} \\
\For{ $node$ in $G$}{
    $k \leftarrow node.QCP$ \\
    $SG_{QCP}[k]$.\textbf{Add($node$)}\\
}
/*\texttt{\footnotesize Stage 2: Find the connected components (i.e., security submodels) for each subgraph from Stage 1 }*/\\
$CC_{QCP} \leftarrow $ \textbf{HashmapInit()} \\
\For{ $QCP$, $SG$ in $SG_{QCP}$}{
    $L_{CC} \leftarrow$ \textbf{BFS($SG$)} \\
    $CC_{QCP}[QCP]$.\textbf{Add($L_{CC}$)}\\
}
/*\texttt{\footnotesize Stage 3: Identify the starting nodes (heads) and end nodes (tails) for each security submodel from Stage 2}*/\\
\For{ $QCP$, $L_{CC}$ in $CC_{QCP}$}{
    \For{ $CC$ in $L_{CC}$}{
    $heads \leftarrow$ \textbf{FindHeads($CC$)}\\ 
    $CC$.\textbf{AddHeads($heads$)}\\
    $tails \leftarrow$ \textbf{FindTails($CC$)} \\
    $CC$.\textbf{AddTails($tails$)}\\
    }

}

$smap \leftarrow CC_{QCP}$

\Return $smap$

\end{algorithm}

\begin{figure}[t]
    \centering
    \includegraphics[width=0.75\linewidth]{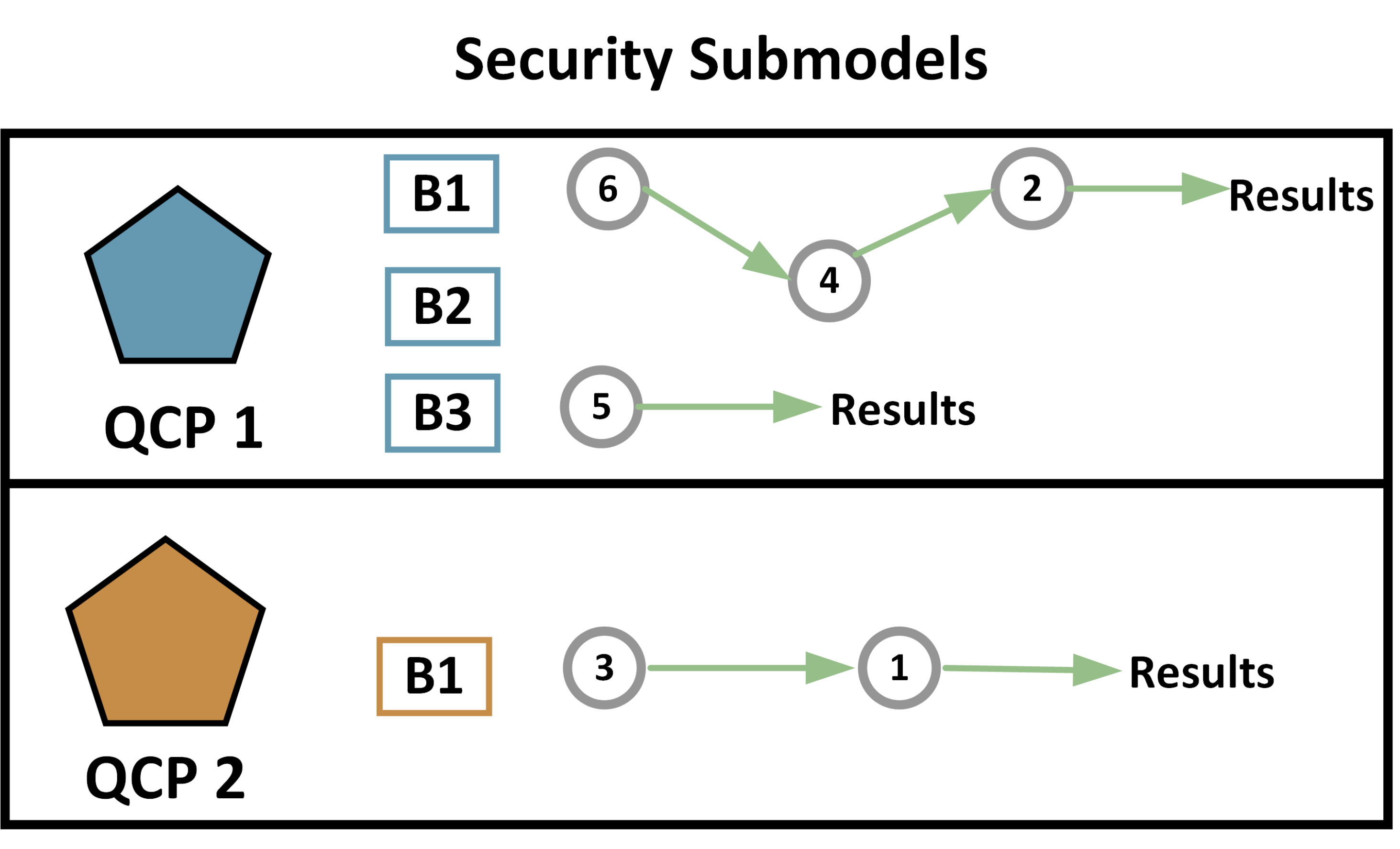}
    \caption{Generation of security submodels of a given QML model in QuMoS for the security measurement}
    \label{fig: subgraph security}
\end{figure}

Fig.~\ref{fig: subgraph security} shows an example of the generation of the security submodels. Based on the system layout in Fig.~\ref{fig: design component} (d) and the node assignment in Fig.~\ref{fig: design component} (f), we can find that node 6, 4, and 2 are connected and all assigned to QCP 1. Therefore, the three nodes form a security submodel for evaluation. Similarly, node 5 is another security submodel in QCP 1, while node 3 and 1 form a security submodel in QCP 2. Note that since no node is assigned to QCP 3, there is no security submodel in QCP 3.

\section{Design of QuMoS Security Engine}
\label{sec:search}










\subsection{Problem Formulation}~\label{subsec:problem definition}
Before describing the design of QuMoS security engine, we formally define the problem as follows: Given the set of quantum cloud providers $P$, the set of quantum neural architecture $A$ and the backbone graph $B$, to design a QML model $M$, the problem is to determine,
\begin{itemize}
    \item[] $\bm{L}$: node-to-node connectivity for all the available computing nodes based on $B$;
    \item[] $\bm{C}$: node-to-architecture mapping for all the available computing nodes based on $A$; 
    \item[] $\bm{D}$: node-to-provider assignment for all the available computing nodes based on $P$;
\end{itemize}
such that $ACC(M)$, the accuracy of QML model $M$ can be maximized while $SecMec(M)$, the security of model is also maximized.


\subsection{Workflow of QuMoS Security Engine}~\label{subsec:security engine}
To find the optimal solution to the problem defined in Section~\ref{subsec:problem definition}, QuMoS proposes a reinforcement learning-based security engine to automatically search for it.

\begin{figure}[t]
    \centering
    \includegraphics[width=1\linewidth]{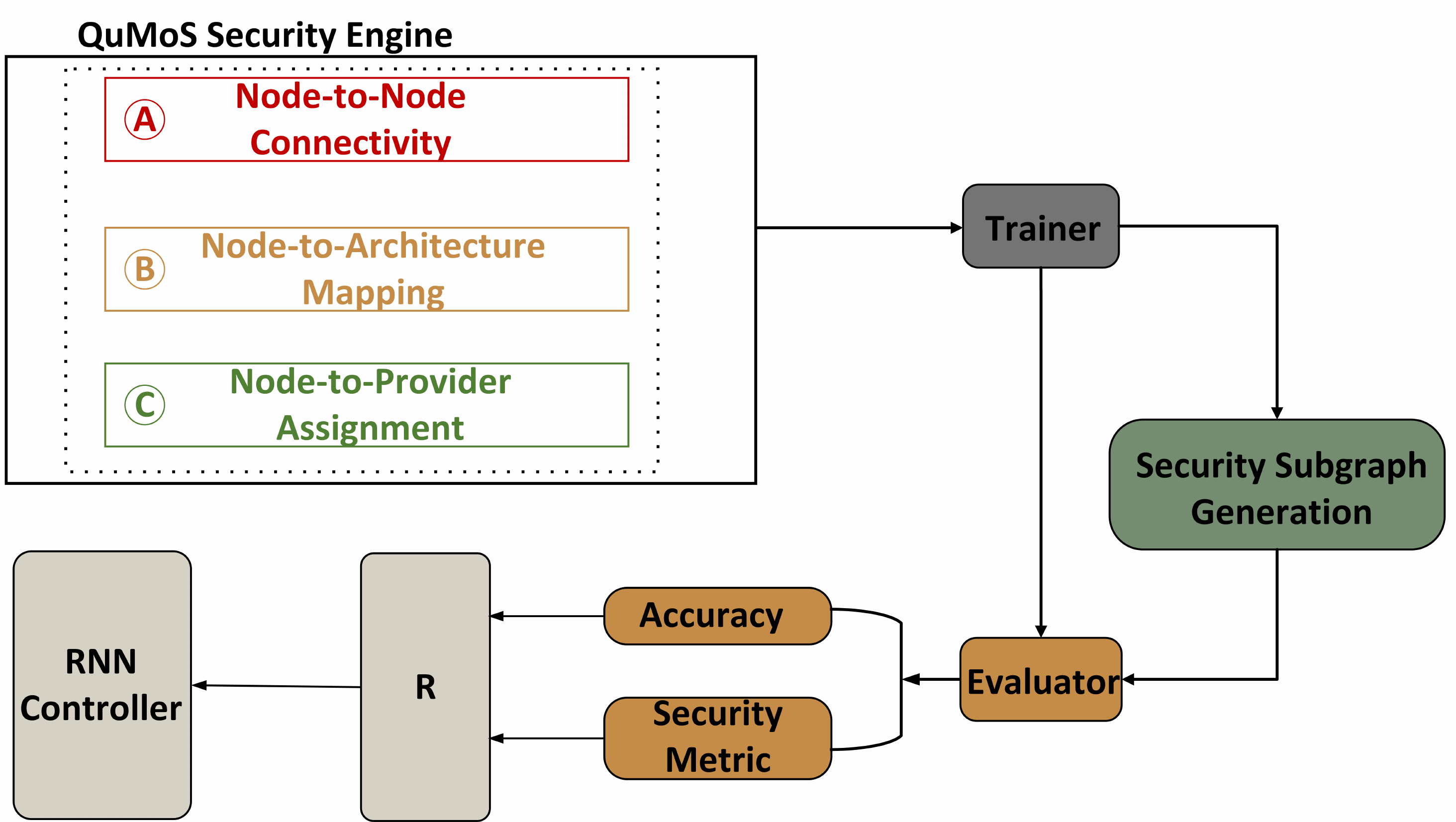}
    \caption{Overview of QuMoS security engine}
    \label{fig: search overview}
\end{figure}

The overview of the security engine is shown in Fig.~\ref{fig: search overview}. The designer needs to provide the backbone graph, the set of quantum neural architecture and the set of quantum cloud providers, which is described in Section~\ref{subsec:properties}, as the input to the security engine. Note that there exists an empty option in the set of quantum neural architecture, which means that the node mapping to this option will be removed from the sampled QML model, together with its related connections. And the input defines the search space for the recurrent neural network (RNN) controller in the security engine. 

As shown in Fig.~\ref{fig: search overview}, given the search space, the RNN controller will sample multiple solutions from it, where each sampled solution will be used to determine the three basic properties of all the computing nodes in a QML model $M$ (i.e., node-to-node connectivity, node-to-architecture mapping and node-to-provider assignment). Each sampled QML model $M$ will then be sent to a trainer to learn its adaptable parameters with train set. Based on the well-tuned model $M$, a set $SMS$ of all the security submodels within $M$ can be generated. The set $SMS$ will then be sent to the evaluator together with $M$ and $QS$ (i.e., the set of all the utilized quantum devices). The evaluator will evaluate its input in terms of accuracy and security. More specifically, the outputs of the evaluator are $ACC(M)$, the accuracy of the sampled model $M$ and $SecMec(M)$, the heuristic security metric of $M$. These two values will then be used to calculate the reward $R$, which is defined as,
\begin{equation}
    R = \frac{1}{|MS|}\sum_{M \in MS} (ACC(M) - b + \lambda SecMec(M))~\label{eq:reward}
\end{equation}

Where $MS$ denotes the set of the sampled QML models from RNN controller. $b$ is the exponential moving average of the previous QML models' accuracies. $\lambda$ is the hyperparameter to control the trade-off between accuracy and security. The calculated $R$ will then be fed to the RNN controller to update its learnable parameters using proximal gradient ascent. This process will be repeated for multiple episodes until the RNN controller converges or achieves the pre-set maximum episodes. 






\section{Experimental Results}
\label{sec:experiment}
\begin{table*}[t]
\centering
\caption{Evaluation of QuMoS on Four Datasets}
\label{tab:Main-Results}
\begin{tabular}{|c|ccccccccc|}
\hline
Dataset & Method & Type & Acc(M) & \begin{tabular}[c]{@{}c@{}}vs. \\ Baseline\end{tabular} & SecMec(M) & \begin{tabular}[c]{@{}c@{}}vs. \\ Baseline\end{tabular} & SecAcc(M) & \begin{tabular}[c]{@{}c@{}}\# Involved \\ QCP\end{tabular} & \begin{tabular}[c]{@{}c@{}}\# Involved \\ Node\end{tabular} \\ \hline
\multirow{7}{*}{MNIST-2} & \multirow{3}{*}{NAS-Single Provider} & QCP 1 & 94.67\% & -0.67\% & 0.0000 & - & 94.67\% & 1 & 7 \\
 &  & QCP 2 & 95.33\% & 0.00\% & 0.0000 & - & 95.33\% & 1 & 5 \\
 &  & QCP 3 & \textbf{95.33\%} & Baseline & 0.0000 & - & 95.33\% & 1 & 7 \\ \cline{2-10} 
 & \multirow{2}{*}{Random Search} & Acc best & 94.67\% & -0.67\% & 0.0246 & -0.4680 & 92.33\% & 2 & 6 \\
 &  & Security best & 90.67\% & -4.67\% & 0.4926 & Baseline & 46.00\% & 3 & 7 \\
 & \multirow{2}{*}{\textbf{QuMoS Security Engine}} & \textbf{Acc best} & \textbf{94.67\%} & \textbf{-0.67\%} & \textbf{0.4401} & \textbf{-0.0525} & \textbf{53.00\%} & \textbf{3} & \textbf{5} \\
 &  & \textbf{Security best} & \textbf{91.33\%} & \textbf{-4.00\%} & \textbf{0.5547} & \textbf{0.0621} & \textbf{40.67\%} & \textbf{3} & \textbf{5} \\ \hline
\multirow{7}{*}{\begin{tabular}[c]{@{}c@{}}Fashion\\ MNIST-2\end{tabular}} & \multirow{3}{*}{NAS-Single Provider} & QCP 1 & \textbf{95.67\%} & Baseline & 0.0000 & - & 95.67\% & 1 & 4 \\
 &  & QCP 2 & 95.33\% & -0.33\% & 0.0000 & - & 95.33\% & 1 & 7 \\
 &  & QCP 3 & 94.67\% & -1.00\% & 0.0000 & - & 94.67\% & 1 & 6 \\ \cline{2-10} 
 & \multirow{2}{*}{Random Search} & Acc best & 92.67\% & -3.00\% & 0.0971 & -0.3624 & 83.67\% & 3 & 7 \\
 &  & Security best & 90.67\% & -5.00\% & 0.4596 & Baseline & 49.00\% & 2 & 3 \\
 & \multirow{2}{*}{\textbf{QuMoS Security Engine}} & \textbf{Acc best} & \textbf{94.00\%} & \textbf{-1.67\%} & \textbf{0.3298} & \textbf{-0.1298} & \textbf{63.00\%} & \textbf{3} & \textbf{7} \\
 &  & \textbf{Security best} & \textbf{91.33\%} & \textbf{-4.33\%} & \textbf{0.4745} & \textbf{0.0149} & \textbf{48.00\%} & \textbf{3} & \textbf{6} \\ \hline
\multirow{7}{*}{MNIST-4} & \multirow{3}{*}{NAS-Single Provider} & QCP 1 & 74.67\% & -1.00\% & 0.0000 & - & 74.67\% & 1 & 4 \\
 &  & QCP 2 & 75.67\% & 0.00\% & 0.0000 & - & 75.67\% & 1 & 7 \\
 &  & QCP 3 & \textbf{75.67\%} & Baseline & 0.0000 & - & 75.67\% & 1 & 6 \\ \cline{2-10} 
 & \multirow{2}{*}{Random Search} & Acc best & 72.67\% & -3.00\% & 0.5046 & -0.1459 & 36.00\% & 3 & 6 \\
 &  & Security best & 68.67\% & -7.00\% & 0.6505 & Baseline & 24.00\% & 2 & 5 \\
 & \multirow{2}{*}{\textbf{QuMoS Security Engine}} & \textbf{Acc best} & \textbf{76.33\%} & \textbf{0.67\%} & \textbf{0.6245} & \textbf{-0.0260} & \textbf{28.67\%} & \textbf{3} & \textbf{5} \\
 &  & \textbf{Security best} & \textbf{72.00\%} & \textbf{-3.67\%} & \textbf{0.7917} & \textbf{0.1412} & \textbf{15.00\%} & \textbf{3} & \textbf{7} \\ \hline
\multirow{7}{*}{\begin{tabular}[c]{@{}c@{}}Fashion\\ MNIST-4\end{tabular}} & \multirow{3}{*}{NAS-Single Provider} & QCP 1 & 79.00\% & -2.67\% & 0.0000 & - & 79.00\% & 1 & 6 \\
 &  & QCP 2 & 80.33\% & -1.33\% & 0.0000 & - & 80.33\% & 1 & 5 \\
 &  & QCP 3 & \textbf{81.67\%} & Baseline & 0.0000 & - & 81.67\% & 1 & 6 \\ \cline{2-10} 
 & \multirow{2}{*}{Random Search} & Acc best & 78.67\% & -3.00\% & 0.4025 & -0.3085 & 47.00\% & 3 & 7 \\
 &  & Security best & 72.67\% & -9.00\% & 0.7110 & Baseline & 21.00\% & 3 & 6 \\
 & \multirow{2}{*}{\textbf{QuMoS Security Engine}} & \textbf{Acc best} & \textbf{80.67\%} & \textbf{-1.00\%} & \textbf{0.5620} & \textbf{-0.1490} & \textbf{35.33\%} & \textbf{3} & \textbf{6} \\
 &  & \textbf{Security best} & \textbf{73.33\%} & \textbf{-8.33\%} & \textbf{0.9409} & \textbf{0.2299} & \textbf{4.33\%} & \textbf{2} & \textbf{3} \\ \hline
\end{tabular}
\end{table*}
\subsection{Experimental Setup}

\textbf{Datasets.} We evaluate QuMoS on MNIST 4-class (0, 3, 6, 9), 2-class (3, 6), Fashion MNIST 4 class (T-shirt, dress, shirt, Ankle boot) and Fashion MNIST 2 class (dress, shirt). All of the images are downsampled to the resolution of $4 \times 4$ and encoded with 4 qubits using the encoder in~\cite{wang2022quantumnas}. And we used the first 300 images in the test set to evaluate the accuracy of the QML models due to the limited quantum computing resources.

\textbf{Design space.} We used a full binary tree with a depth of 3 to serve as the backbone graph in the experiments, where the total number of computing nodes is $7$. There are 6 optional quantum neural architectures in the set of quantum neural architectures, including 1 empty option and 5 neural architecture design  from~\cite{sim2019expressibility}. For the set of quantum cloud providers, there are three options. Due to the limited access to multiple providers, we used the three noisy simulators $ibmq\_quito, ibmq\_belem$ and $ibmq\_manila$ in IBM Quantum to simulate the three cloud quantum providers, where each of the providers only has one quantum device. And we use QCP 1, QCP 2, and QCP 3 respectively, to refer to these three devices in the experiments. Note that the size of the search space is $(6\times 3)^{7} \approx 10^{9}$, which is huge.

\textbf{Setting of security engine.} The implementation of the security engine is based on TorchQuantum~\cite{wang2022quantumnas,wang2022quest}, IBM Qiskit and Tensorflow. The RNN controller utilizes Long short-term memory (LSTM) cells, where each has 35 hidden units. It is trained for 200 episodes and it samples 1 solution at each episode. It is optimized by RMSProp algorithm with a decay rate of 0.9, while the initial learning rate for the optimizer is set to 0.99. Besides, for the value of $\lambda$ in the reward function, we tried 0.1, 0.5 and 1 to find the best trade-off between accuracy and security. For each sampled solution, it is trained for 20 epochs with a batch size of 64. We use Adam optimizer with an initial learning rate of $5 \times 10^{-3}$ and weight decay of $10^{-4}$. The learning rate is gradually tuned by a cosine learning rate scheduler. The quantum circuit of the well-tuned sampled solution is run with 8092 shots and the optimization level for its compilation is set to 2.

\subsection{Main Results}
Table~\ref{tab:Main-Results} shows the main results of our experiments to evaluate QuMoS on (Fashion) MNIST-2 and (Fashion) MNIST-4. Two types of methods are proposed for comparison. NAS-single provider corresponds to the method which utilizes the reinforcement learning-based neural architecture search (NAS) like the QuMoS security engine but without considering the assignment to multiple quantum cloud providers in the design space. It means that the generated solution $M$ is executed on a single cloud provider,  and thus no security guarantees are provided for the solutions of this type of method, i.e., $SecMec(M) = 0$. $SecAcc(M)$ corresponds to the term $\max_{q \in QS, sm \in SMS} ACC(sm, q)$ in Equation~\ref{eq:secmec}, which refers to the maximum accuracy of all the generated security submodels contained by a given QML model $M$. $SecAcc(M)$ is thus a more intuitive proxy metric to show the security of $M$ and a larger value of $SecAcc(M)$ usually implies lower security of $M$. Therefore, for solutions from NAS-single provider, we have $SecMec(M) = Acc(M)$. Note that in Table~\ref{tab:Main-Results}, we reported the solution with the best accuracy on any of the three available quantum cloud providers separately. Moreover, the maximum $Acc(M)$ of the three NAS-single providers will serve as the baseline for the comparison of $Acc(M)$. 

The other baseline is the random search, which shares the same design space as QuMoS security engine but explores it randomly instead of being guided by an RNN controller through reinforcement learning. For a fair comparison, the number of sampled solutions of random search is the same as that of QuMoS security engine. 

Two types of solutions for the random search and QuMoS security engine are shown in Table~\ref{tab:Main-Results}, i.e., the solution with the best accuracy and the solution with the best security. Besides, the $SecMec(M)$ of the solution with the best security in the random search serves as the baseline for the comparison of $SecMec(M)$.


\begin{figure}[t]
    \centering
    \includegraphics[width=3 in]{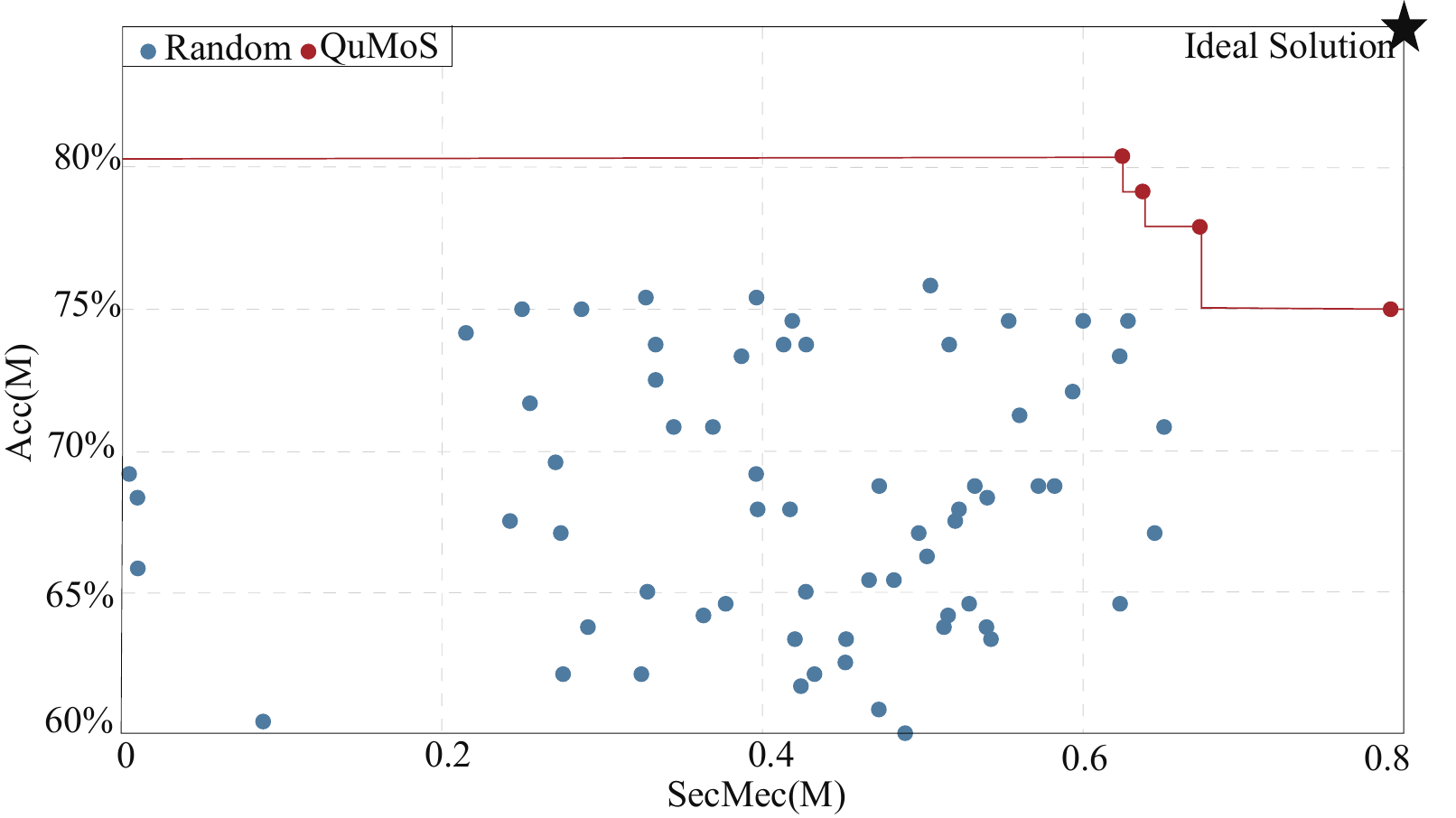}
    \caption{Pareto frontier of QuMoS dominates random solutions on MNIST-4}
    \label{fig: pareto optimal}
\end{figure}


For the solutions with the best accuracy, it is obvious that the solutions of QuMoS security engine outperform the solutions of random search consistently on all of the evaluated datasets in terms of $Acc(M)$. And the maximum improvement on $Acc(M)$ over random search can achieve $3.67 \%$ on MNIST-4, which is even $0.67\%$ higher than that of the best NAS-single provider (i.e., the indicated baseline in Table~\ref{tab:Main-Results}). Besides, on MNIST-2, Fashion MNIST-2 and Fashion MNIST-4, the solutions of QuMoS security engine are also close to the baseline, with a maximum accuracy loss of $1.67 \%$ on Fashion MNIST-2. Moreover, the solutions of QuMoS security engine are also more secure than the solutions from random search. The maximum improvement on $SecMec(M)$ over random search can achieve about $0.42$ on MNIST-2 while the minimum improvement is about $0.12$ on MNIST-4. These results demonstrate that QuMoS security engine can indeed make a better trade-off between accuracy and security than the random search.  

Similarly, for the solutions with the best security, QuMoS security engine can still beat the random search consistently on all of the evaluated datasets in terms of security. And the maximum improvement on $SecMec(M)$ over random search can achieve about $0.23$ on Fashion MNIST-4, while the minimum improvement on $SecMec(M)$ over random search is about $0.06$ on MNIST-2. Besides, the solutions of QuMoS security engine are also more accurate than that of the random search. The maximum improvement on $Acc(M)$ over random search can achieve about $3.33 \%$ on MNIST-4 while the minimum improvement is about $0.67 \%$ on Fashion MNIST-4.

In Table~\ref{tab:Main-Results}, the column \# Involved QCP denotes the number of quantum cloud providers which are actually utilized by the reported solutions. Although most of the solutions of QuMoS security engine make full use the of three available quantum cloud providers, there is an exception for the solution on Fashion MNIST-4, it only uses two quantum cloud providers to achieve the best security. This result indicates that QuMoS security engine indeed provides the flexibility to allow the QML model to skip some optional cloud providers for best performance.  

In Table~\ref{tab:Main-Results}, the column \# Involved Node denotes the number of computing nodes existing in the reported solutions. Although the backbone graph provides 7 nodes for selection, it clearly shows that to achieve the best performance, most of the solutions do not need to use all of the optional nodes. It also proves that QuMoS security engine provides flexibility in node-to-node connectivity, which allows the deactivation of some computing nodes in the backbone graph and their related connections.

\subsection{Search Space Exploration}

Fig.~\ref{fig: pareto optimal} reports the search space exploration for QuMoS security engine and the random search on MNIST-4, where the y-axis is $Acc(M)$, the accuracy of the sampled QML model M, and the x-axis is $SecMec(M)$, the security of the sampled QML model M. The ideal solution is located in the right-top corner in Fig.~\ref{fig: pareto optimal}, denoted as a star. 

In Fig.~\ref{fig: pareto optimal}, the red points correspond to the solutions lying on the Pareto frontier for the QuMoS security engine, while the blue points denote the solutions generated by the random search. It is obvious that all of the solutions from random search are under the Pareto frontier generated by the QuMoS security engine, which means that the QuMoS security engine can consistently push forward the Pareto frontier compared with random search.

\subsection{Importance of Model
Design: A Case Study}~\label{subsec:motivation}
\begin{figure}[t]
    \centering
    \includegraphics[width=3in]{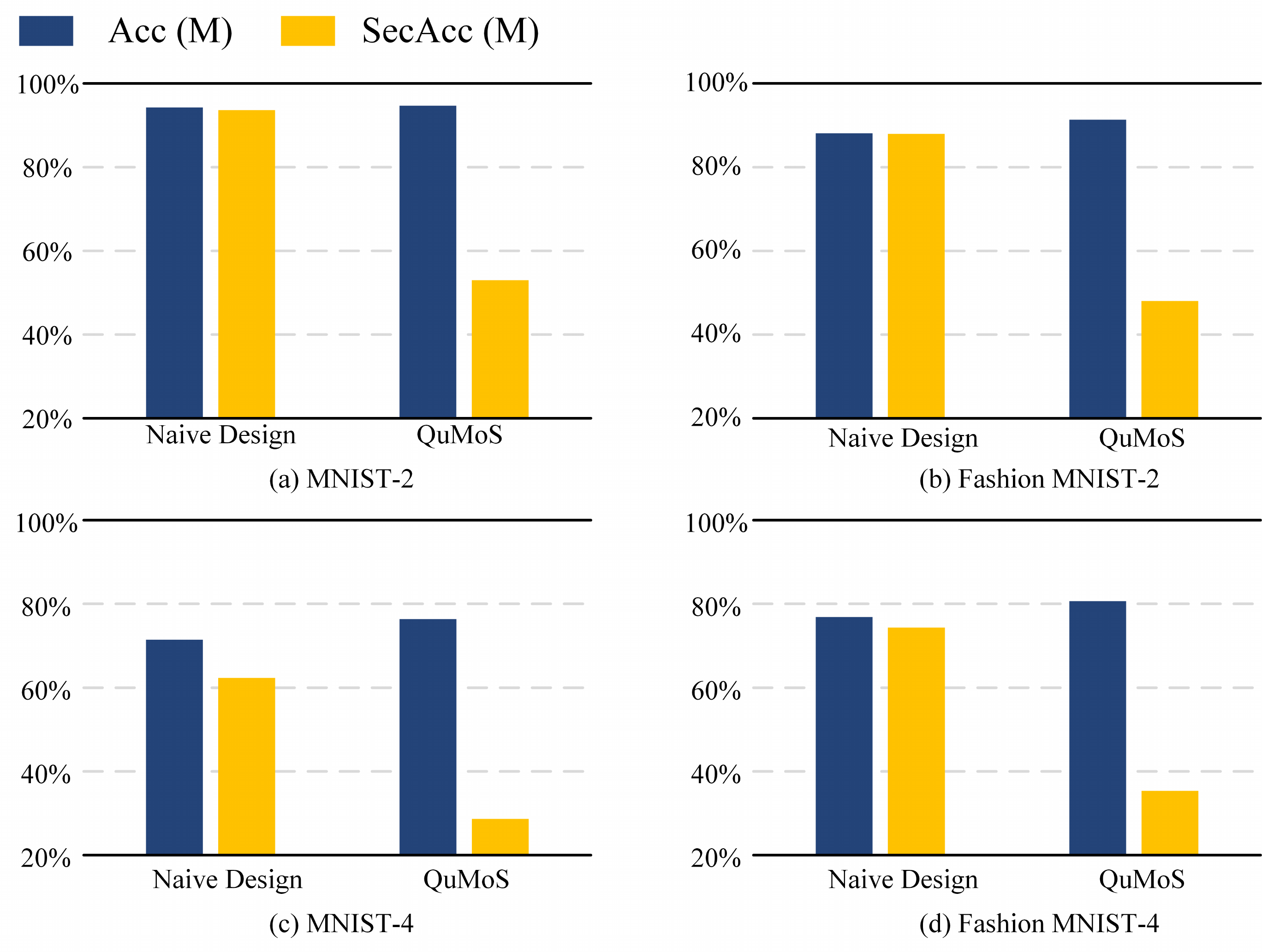}
    \caption{Comparison between the model design from QuMoS and the naive design }
    \label{fig: motivation}
\end{figure}

Fig.~\ref{fig: motivation} shows a case study on four datasets to highlight the importance of the design proposed by QuMoS in Section~\ref{sec:design}. The naive design is manually crafted and has 5 computing nodes with the same quantum neural architectures. And for QCP-1, and QCP-2, they are assigned 2 nodes while for QCP-3, it is assigned 1 node. The y-axis of the blue bars is $Acc(M)$ while for the yellow bars, it is $SecAcc(M)$. When $Acc(M)$ is close to $SecAcc(M)$, it means that the partial model of $M$ obtained in a single quantum cloud provider can achieve a close accuracy to the complete model $M$, which indicates low security. 

In Fig.~\ref{fig: motivation}, it is obvious that the model with naive design has low security on all the evaluated datasets, while for the model from QuMoS, the security is high and the accuracy is also improved. Therefore, we can conclude that the design of the QML model under distributed setting is a non-trivial task. And QuMoS can indeed find the design achieving a good trade-off between security and accuracy.

\subsection{Robustness Analysis of QuMoS}

\begin{figure}[t]
    \centering
    \includegraphics[width=3.5 in]{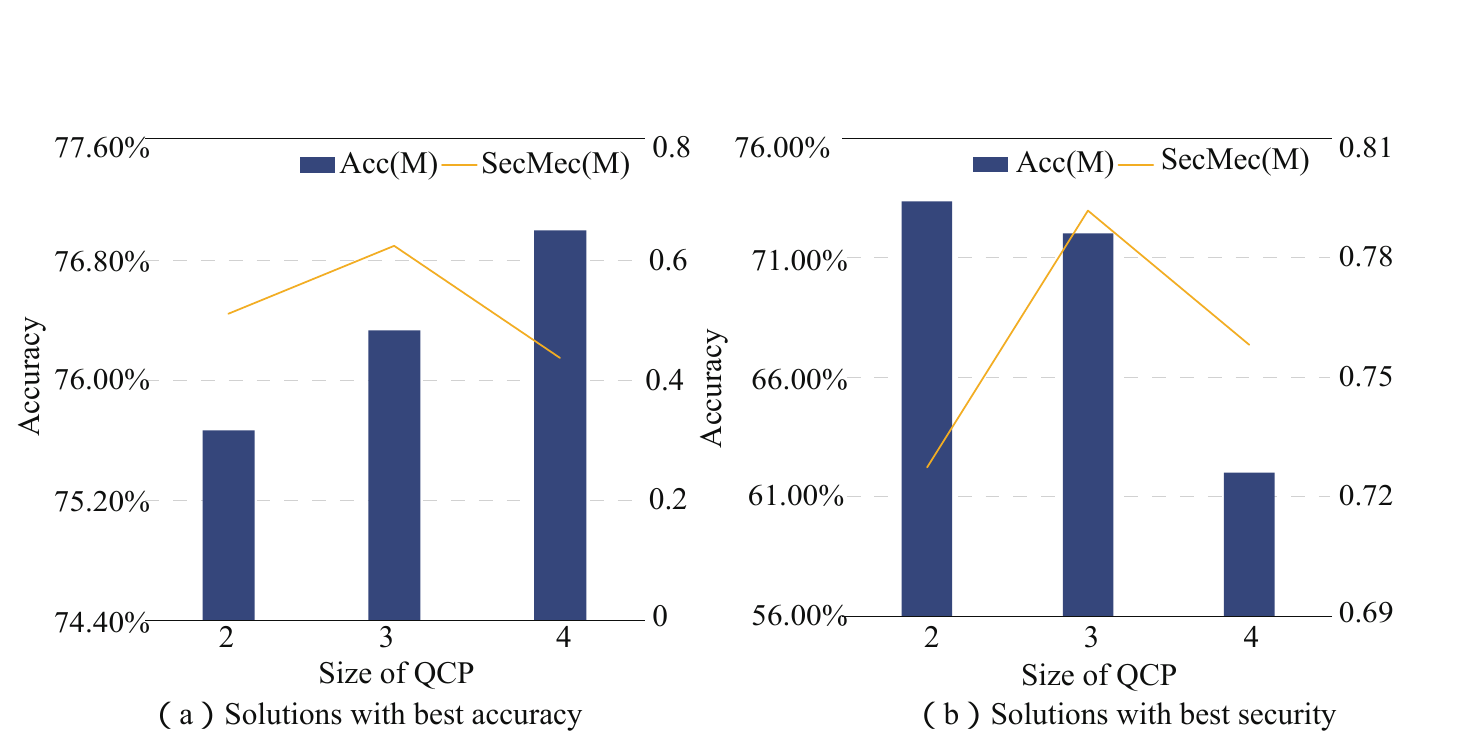}
    \caption{Robustness analysis of QuMoS to the size of QCP}
    \label{fig: robustness}
\end{figure}

We conducted experiments to explore the robustness of QuMoS to the size of the set of quantum cloud providers on MNIST-4. The experimental results are shown in Fig.~\ref{fig: robustness}, where (a) shows the solutions with the best accuracy while (b) shows the solutions with the best security. In Fig.~\ref{fig: robustness}, the y-axis is $Acc(M)$ for the blue bars and it is $SecMec(M)$ for the orange lines. The x-axis represents the size of the set of quantum cloud providers. More specifically, we eliminated $ibmq\_belem$ when the size is 2 and added $ibmq\_lima$ when the size is 4. For a fair comparison, we set the number of episodes to 200 for all the settings. 

Based on the results, we can conclude that QuMoS can work even when the QML service provider only accesses 2 quantum cloud providers. For example, in the case of the solution with the best security, there is a $1.33 \%$ accuracy improvement and a decrease of $0.064$ on security compared with the counterpart when the size is 3, providing a similar level of trade-off between accuracy and security. When the size is 4, the security of QML suffers from a little degradation. It implies that more episodes are needed for the search algorithm since the search space has been expanded by about 10X.





\section{Related Work}\label{sec:related work}

\textbf{Quantum Machine Learning (QML)}. Quantum machine learning focuses on the design of machine learning algorithms suitable for the execution on quantum computers.  And the most popular QML algorithm is based on variational quantum circuit (VQC)~\cite{sim2019expressibility,wang2022quantumnat,zheng2022benchmarking,liang2022variational,wang2022qoc, schuld2020circuit,schuld2021effect,romero2021variational,chen2020variational,schuld2019quantum, hu2022quantum}, which utilizes the unique properties of entanglement in quantum computing to perform machine learning tasks. Recently, another type of QML model~\cite{tacchino2019artificial,jiang2021co, wang2021exploration, liang2021can, tacchino2019artificial, jiang2021machine} is emerging, which follows the idea to mimic the neuron computation in classical deep neural networks. In this paper, we focus on the QML model based on VQC. 

\textbf{Security in Machine Learning}. Secure computation can be used to protect the machine learning process by performing computations on encrypted data, thereby ensuring that outsourced computation and data remain secure in the cloud. Based on this idea, a variety of techniques are proposed, including the methods based on homomorphic encryption~\cite{gilad2016cryptonets}, secure multiparty computation~\cite{riazi2018chameleon, peng2023rrnet} and the work integrating the two approaches~\cite{juvekar2018gazelle}. However, these methods primarily focus on classical computing environments and may not be applicable in quantum computing scenarios, as encryption methodologies that work well in classical computing may not be effective in quantum computing.

\textbf{Security in Quantum Computing}. Blind quantum computing~\cite{fitzsimons2017private,broadbent2009universal} has been studied for years to preserve the security of quantum circuits. However, this type of technique assumes that the local user has a local quantum terminal that could rotate and measure a single qubit, which does not hold in our cloud-based scenario.~\cite{huang2017experimental} proposes a distributed algorithm to protect the quantum circuit when the local user only has a classical terminal and communicate through the quantum network~\cite{yurke1984quantum,zhan2022transmitter}. However, this algorithm is customized to the application of prime factorization and assumes that the distributed servers could share entanglement. Therefore, it is not applicable to QML applications.~\cite{mahadev2020classical} proposes an encryption algorithm to protect the data for quantum computing, while our paper focuses on protecting the computing models.




\section{Conclusion}\label{sec:conclusion}
Quantum machine learning is an emerging application of quantum computing. Since the near-term quantum computers are usually remotely accessed through the cloud, a well-tuned QML model developed by the company is delegated to a quantum cloud provider as a service to ordinary users. In this case, the QML model will be leaked if the cloud provider is under attack. Therefore, we propose QuMoS, which could automatically generate optimized strategies to distribute the QML model to multiple providers, where each provider only has the partial model with bad performance while maintaining the high performance of the complete model. 

\section*{Acknowledgment}
This work is supported in part by Mason’s Office of Research Innovation and Economic Impact (ORIEI), Quantum Science and Engineering Center (QSEC) and Center for Trusted, Accelerated, and Secure Computing \& Communication. The research used IBM Quantum resources via Los Alamos National Lab Hub. And this project was supported by resources provided by the Office of Research Computing at George Mason University (URL: https://orc.gmu.edu) and funded in part by grants from the National Science Foundation (Awards Number 1625039 and 2018631). In addition, the research used resources from the Oak Ridge Leadership Computing Facility at the Oak Ridge National Laboratory, which is supported by the Office of Science of the U.S. Department of Energy under Contract No. DE-AC05-00OR22725.
We also acknowledge TorchQuantum to support this work.



\bibliographystyle{IEEEtranS}
\bibliography{reference,quantum}

\end{document}